\documentclass[runningheads]{llncs}

\usepackage[pagewise]{lineno}

\usepackage{hyperref}
\usepackage{verbatim}
\usepackage{graphicx}
\usepackage{color,soul}
\usepackage{adjustbox}
\usepackage{adjustbox,lipsum}
\usepackage{pgfplots}
\pgfplotsset{width=7cm,compat=1.8}
\usepackage{pgfplotstable}

\usepackage{amsmath}

\title{Detection and Analysis of Drive-by Downloads and Malicious Websites}

\author{ Saeed Ibrahim\inst{1} \and Nawwaf Al Herami\inst{1} \and Ebrahim Al Naqbi\inst{1}  \and Monther Aldwairi \inst{1}}
\authorrunning{Saeed Ibrahim et al.}
\institute{College of Technological Innovation\\
	Zayed University,
	Abu Dhabi, UAE 144534\\
  \email{\{m80007514,m80006805,m80006809,monther.aldwairi\}@zu.ac.ae}
}

\begin{document}

\maketitle

\begin{abstract}
A drive-by download is a download that occurs without user’s action or knowledge. It usually triggers an exploit of vulnerability in a browser to downloads an unknown file. The malicious program in the downloaded file installs itself on the victim’s machine. Moreover, the downloaded file can be camouflaged as an installer that would further install malicious software. Drive-by downloads is a very good example of the exponential increase in malicious activity over the Internet and how it affects the daily use of the web.
In this paper, we try to address the problem caused by drive-by downloads from different standpoints. We provide in-depth understanding of the difficulties in dealing with drive-by downloads and suggest appropriate solutions. We propose machine learning and feature selection solutions to remedy the drive-by download problem. Experimental results reported 98.2\% precision, 98.2\% F-Measure and 97.2\% ROC area.

\vspace{6pt}\textbf{Keywords:} drive by download, malware detection, web security
\end{abstract}

\section{Introduction}

Miscreants make use of malicious web content to perform attacks targeting web clients. Drive-by downloads (DBD) are unintentional downloads of malware or virus on to a mobile device or a computer. Due to the increased population of several web applications, DBD have become one of the most common malware spreading methods, thereby leading the security threats to cyber community. According to \cite{mavrommatis2008all}, query search results from Google contain more than 1.3\% of the web pages that do DBD attacks. These downloads are located on normal-looking, but malicious websites \cite{aldwairi2011malurls}. They exploit vulnerabilities in out-of-date apps, browsers, plugins, or operating systems. Over the years, hackers have become much more sophisticated that just opening such web page could allow malicious code to be installed on the device without the knowledge and consent of the user. Downloaded malware takes complete control of the victim’s platform \cite{cova2010detection}. Once the attacker gets full control, he can download and execute any code and run malicious activities on the victim's platform such as joining botnets, sending spam emails, and participating in distributed denial of service attacks \cite{egele2009defending}. Attackers may also record keystrokes, steal passwords, and can access sensitive information. Use of DBD to steal confidential data is also a major threat to the financial companies and banks.

A DBD attack occurs in four steps. First, the attacker compromises a genuine website and uploads malicious content to it. When a user visits that website, the malicious program is downloaded by browser, installed by itself, and the attacker gets full control \cite{egele2009defending}. APT programs and methods used by cybercriminal groups to attack businesses make them more dangerous.

In 2015, almost two million cases of malware infections to steal money were registered, while 34.2 \% of computer users were exposed to at least one such attack through the year \cite{anton2015kaspersky}. In order to ensure protection against such attack, there is a vital need for new methods and technologies that can safeguard the users from DBD attacks \cite{vergelis2015kaspersky}. There are couple of existing techniques to detect and prevent such attacks. The detection of attacks can be performed by tracking web addresses with a history of malicious behavior \cite{mavrommatis2008all}. According to Microsoft, Bing normally detects huge numbers of DBD pages every month. However, after getting blocked by Bing, the attackers switch servers and thus the same attacks are reborn but with different domain names \cite{zhang2011arrow}. Intrusion detection systems monitor traffic and system activities  and may be used to detect attacks \cite{aldwairi2017detection}.

In order to counter the innovative tactics employed by the hackers, there is a vital need to develop efficient techniques that could potentially counter DBD attacks. In this paper, we proposed a novel design, which uses machine learning to detect and prevent DBD attacks. We selected nine attributes from a dataset of benign URLs from University of California Irvine (UCI) machine learning repository and malicious URLs from malware domain list \cite{MALURLS2012}. Each attribute was chosen carefully to measure its effectiveness on different characteristics of malicious URLs. Furthermore, we employed several machine learning models for the training the system to detect malicious URLs. However, after empirical performance evaluation of these models, we selected Naive Bayes (NB), JRip, and J48 classifiers.

The rest of the paper is structured as follows: Section \ref{sec-2} contains the related work. Section \ref{sec-3} describes the methodology. Results are discussed in section \ref{sec-4} and section \ref{sec-5} concludes our work.

\section{Related Work}
\label{sec-2}
Below we classify the most relevant work on detecting DBDs.

\subsection{Using web crawler to detect drive by downloads} 
Harley and Pierre-Marc work does not offer a solution to DBDs, but tries to provoke more research in the area by suggesting possible ideas \cite{harley2008drive}. It provokes researchers to pay more attention to attacks that are large scale in nature and which do not use codes that are self-propagating. This is because current attacks are sophisticated and, therefore, a long-lasting solution may be one that uses the fault-tolerant and robust software in addition to ensuring the monitoring of web pages. A web crawler can be used to identify distribution points, however, due to the complexity of this detection, false positives risks can be lessened by either digital signing or obfuscating techniques can be avoided. Some of the characteristics of this web crawler would be; ability to analyze HTML pages as well as follows its links; ability to imitate web cookies; ability to imitate scripting languages in order to decode obfuscated code; and ability to use heuristics in the detection of possible exploits in web pages. It concludes that measures, which are semi-effective and multi-layered, and those that accept specific risks of both false positives and negatives offer much protection.

\subsection{Antivirus software to detect drive-by downloads malware}

Narvaez et al., studied how antivirus software can be useful in the detection of drive-by malware installation by studying the effectiveness of the current antivirus tools \cite{narvaez2010drive}. A sample of malware was collected by use of a honeypot. The sample of the malware was categorized into whether the malware used either delivered payload or downloader. An evaluation of the results was made by common antivirus software to determine their effectiveness in detecting exploits. After 30days, the sample of the malware was scanned again as it was expected that the antivirus would have made an update of signature databases. According to the initial results, Norton detected 66\% of the collected malware, Kaspersky 91\%, CA 61\%, ClamWin 62\% \cite{Flaifel2018WuManber} and TrendMicro 69\% \cite{7921994}. The next scan, after 30days,  showed an increase in the rate of detection with Norton having 90\%, Kaspersky 98\%, TrendMicro 70\%, ClamWin 75\% and CA 81\%. However, even though there was an improvement in the second scan, signature-based antivirus may not perform well in reality. This is because just as they had an opportunity to perform an update on their signatures similarly would attackers update malware. The initial detection, which was low, shows that malware authors use polymorphic capabilities. In 84\% of attacks, downloaders are used instead of payloads. Antivirus products struggle to keep their signature databases up to date with the continuously changing threat landscape \cite{7894031}.

\subsection{BrowserGuard as a behavior-based solution} Hsu et al. \cite{hsu2011browserguard} proposed a behavior-based BrowserGuard, which detects secret downloads and blocks the malware from being executed. BrowserGuard uses two phases to provide protection to its host. The first is the filtration phase, whereby BrowserGuard makes a distinction between malicious and benign files depending on the situations in which they are downloaded. The second is the prohibition phase, whereby a request for the execution of malicious files is denied. In order to test the technique in terms of false positives, BrowserGuard visited the 500 top-ranked websites from Alexa. As expected BrowserGuard did not issue any attack alert, therefore, BrowserGuard had zero false positives. To measure the false negatives, Metasploit framework  was used to generate ten malicious web pages that are then hosted on a remote server. BrowserGuard blocked all ten pages, therefore, the authors claimed zero false negatives. To assess the performance overhead of BrowserGuard, the time to download fives web page, from Alexa, was measured 2000 times. BrowserGuard introduced a fixed delays time and the worst performance overhead was 2.5\%. Unfortunately, we believe the test samples are insufficient to support the conclusions and BrowserGuard only works for Windows Internet Explorer 7.0.

\subsection{A framework for DBD attacks with users voluntary monitoring of the web} 
Matsunaka et al. \cite{matsunaka2013detecting} proposed participative monitoring framework that fights DBDs with voluntary monitoring of websites by users and expert analysts. The framework provided a security ecosystem whereby users allow monitoring of their web activities, while security analysts do an inspection of the information in order to detect threats, devise countermeasures and provide feedback to the users. The framework enables users to provide data via the sensors and security analysts to give feedback through analyzing the data available at the center. The sensors are located in web proxies, DNS servers, and web browsers. Additionally, a web crawler was used to inspect web pages that are suspicious. The real-time data enabled the framework to previously detect unknown malicious web pages. However, advertisement hosts can cause false positives and further work is needed to address that.

\subsection{HTML and JavaScript feature for detecting the drive-by download} 
Priya et al. \cite{priya2013static} provided a static approach to the detect DBDs using JavaScript and HTML features \cite{priya2013static}. A sample dataset was created with 311 malicious URLs, from www.malwaredomainlist.com, and 654 benign URLs from Alexa were used to test different classifiers. To view the source code of benign sites you just open the URL, however, opening a malicious web page is a problem because it will cause malware to be installed on the computer. Therefore, MATLAB parser was developed to extract the malicious source code without visiting and executing the code. The HTML code was parsed and JavaScript and HTML features were extracted.They used both WEKA and MATLAB to evaluate the classifiers performance with 92\% best case detection accuracy.

\subsection{Approach to detect drive-by download based on characters} 
Matsunaka et al. \cite{matsunaka2014approach} proposed FCDBD that includes monitoring sensors on the client side and analysis center on the network. The sensors include web browsers, web sensors or DNS sensors. The browser sensors extracted the user's data while DNS and web sensors monitored DNS-/HTTP- related traffic \cite{Switch}. The analysis center collects the logs and analyzes them, if malicious websites are detected, the information is reported to monitoring sensors so the users may not access the websites. The approach was evaluated using D3M 2013 dataset. According to the results, false positives only occur when a transition of a sequence of web pages is terminated before the malware is downloaded. To compensate for that, advertisement or affiliates scripts are obfuscated and referrer field is empty.

\subsection{Enhanced approach for malware downloading: } 
Adachi et al. \cite{adachi2015approach} used two approaches to predict DBD through opcode and vulnerability evaluation. The first approach identified vulnerabilities CVE-IDs in the web pages to predict the of malware download. For analysis, Wepawet was employed to identify CVE-IDs in the web pages, and the National Vulnerability Database (NVD) provided information concerning the CVEs. To improve detection rates they are reduced unnecessary information by a grouping algorithm \cite{Mohammad_2019}. Features were then extracted and the prediction model computed malware-downloading probabilities. The second approach combined opcode with the first approach one because opcode by itself fails to detect attacks that do not use JavaScript. Pages from 2011-2014 D3M datasets and AlexaTop500 were used. The first approach had 83\% prediction accuracy and low FPs rate, however it had high FNs rate. The second approach had a 92\% prediction accuracy, 11\% FNs and 6\% FPs using Random Forest.

\subsection{Analyze redirection code for mining URLs:} Takata et al. \cite{takata2015minespider} MineSpider performed an analysis on JavaScripts that include browser fingerprinting and redirection code and extracted possible URLs through the execution of the redirection code. MineSpider applied program slicing to JavaScript in order to extract execution paths, the extracted code fragments are executed by an interpreter and URLs are extracted. The outcome is just URL extraction and no detection was done. However, the URLs extracted by this method can be analyzed for malice using other approaches. MineSpider could extract more than 30,000 URLs in seconds compared to other methods.

\subsection{Visualize the flow of HTTP traffic} 
Kikuchi et al. \cite{kikuchi2015automated} used decision trees to classify DBDs by using features such as object size and redirection methods. The first premise was that many code variations modify words that are user-defined without the structure of the script being affected. Second, the characteristics of the scripts do not protect from DBDs because of disguised transformations fabrication. Additionally, they used the prediction of latent behavior to detect large-scale DBDs by using the  drive-by disclosure method, which bridges the gap in between static and dynamic approaches. The method captured models and learned latent behaviors as opposed to scanning web pages for content that is malicious. To evaluate the efficiency of the approach 50 malicious and 50 legitimate sessions were obtained from Alexa. It was found that the method had no false positives but had 0.06 chance of false negatives. The results showed that drive-by exposure can filter out scripts that are benign in nature, detect malicious scripts, and detect a variety of obfuscated patterns of DBDs as well as sort-out scripts that are disguised. In comparison to other high-tech solutions, drive-by disclosure was doubling accurate when compared to Cujo and it outdid JSAND by 29\%.

\subsection{Drive-by download as a large scale web attacks }
Jodavi et al. drive-by disclosure \cite{al2015drive} used anomaly DbD hunter approach to train and detect using a collection of classifiers. In the training stage, inputs of benign web pages are run in a browser. Then, JavaScript byte codes are logged for the web pages and a feature vector generated for the sequence. The feature vectors are then used to construct the classifiers baseline. The detection stage involved logging JavaScript byte codes for web pages, after which a feature vector is generated and applied to all base classifiers. The detection performance of DbD hunter was evaluated and was found that it increased the rate of detection by 12.44\%, while decreasing rates of false alarms by approximately 48.13\%. It had an accuracy of 97\%, a detection rate of 96.3\% and false alarm rate of 1.8\%.
Anomaly detection approach \cite{IRECAP20971} have been used to detect DBD. According to \cite{jodavi2015dbdhunter}, attacks by DBDs make use of browser exploit packs (BEPs) that are deployed on compromised servers to spread malware. BEPs that are widely used include sweet orange, Black Hole, Angler, Nuclear, Sakura, Fiesta, Hunter, Magnitude and Styx. The study makes an analysis of features that are built-in, which allow successful attacks by DBDs. The study conclude that just as attacks by DBDs increase in sophistication, so should the solutions.

\section {Methodology}
\label{sec-3}
We develop a novel mechanism to counter DBD attacks that employs machine learning techniques. The proposed mechanism is able to classify the URLs into benign and malicious categories accurately.The benign category refers to websites that are safe, whereas the malicious category relates to the websites created by attackers to gain access or retrieve sensitive information.  We used Waikato Environment for Knowledge Analysis (WEKA) \cite{holmes1994weka} to classify the URLs based on different attributes using machine learning based models. WEKA is a popular machine learning suite developed at the University of Waikato, New Zealand and is licensed under the GNU General Public License (GPL). It contains machine learning algorithms for data mining related tasks. Integration feature helps to integrate these algorithms with the application code. It also supports data pre-processing, classification, regression, clustering, association rules, and visualization. The following subsection summarize the methodology used to classify DBDs and evaluate the performance.

\subsection{dataset}
We collected benign URLs from open source UCI Machine Learning Repository \cite{Lichman:2013} and we used a list of 63 updated malware and spyware URLs from Malware Domain List \cite{MDL}.

\subsection {Feature selection}
Feature selection, also known as variable selection or attribute selection, is a process to select relevant features from predictive models. Each instance of the dataset used by machine learning algorithms is represented by the same set of features. These features can be continuous, categorical, or binary. We selected multiple effective features to build our proposed model. Given a single URL, its features were extracted and categorized into eight attributes (plus class) that were used by WEKA as itemized below.

\begin{itemize}
\item HostRank: the URL’s global Amazon Alexa ranking \cite{alexa}.
\item CountryRank: the URL’s Amazon Alexa website rank by country \cite{alexac}.
\item ASNNumber: The autonomous system number (ASN), which is assigned to the URL’s domain, and used in BGP routing. \cite{huston2006exploring}.
\item DotsInURL: number of dots in URLs \cite{ALDWAIRI2018215}.
\item Lenghthofurl: length of the URL.
\item IPaddresss: is the host name using ip address rather than name address.
\item Lengthofhostname: length of host name.
\item Safe Browsing: rating of Google safe browsing.
\end{itemize}

\begin{figure}
	\centering
	\includegraphics[width=.7\textwidth]{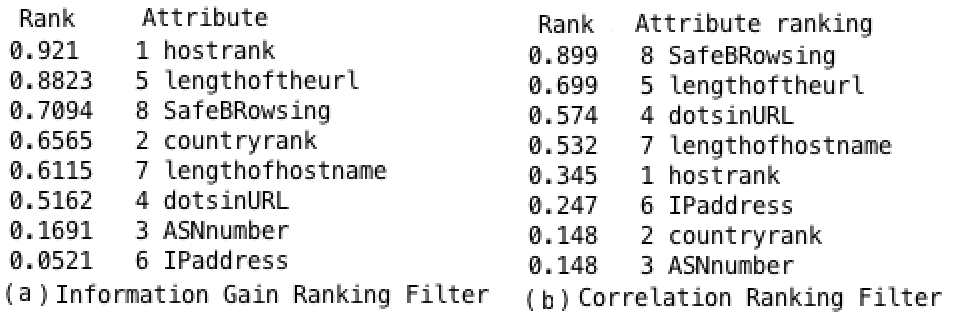}
	\caption{Information Gain and Correlation Ranking Attributes Ranking}
	\label{fig:afigure}
\end{figure}

Two attribute evaluators: Correlation Attributes Evaluation (CAE) and Information Gain Attributes (IG) have been used on the dataset. Correlation Attributes Evaluation is used to choose best attributes for model training. It measures the correlation between attribute and the class and evaluates its worth. Information Gain picks attributes by measuring IG with respect to the class. For this work, eight features were selected to be used with WEKA. Referring to the figure \ref{fig:afigure}, most of the attributes have scored a high ranking except IPaddress and ASNnumber for which, IG  was 0.0521 and 0.1691, respectively. On the CAE, the IPaddress and ASNnumber scored 0.247 and 0.148, which are the lowest scores in the precision test. Thus, these two attributes were eliminated from the attribute set. We finalized six features that include Host Rank, Country Rank, Dots in URL, Length of the URL, length of the host name, in addition to the class: malicious or benign.

\subsection{Classification}
 Many classifiers were chosen to train on the selected dataset, however, NB, JRip, and J48 outperformed all others. Therefore, we experimentally determined that those three are the best classifiers based on their performance on a given dataset. To evaluate the trained model, we employed 10 folds cross validation. Cross-validation is a technique to evaluate predictive models by splitting the original dataset sample into a training and test sets to train and evaluate the model respectively. The process is repeated $k$ times, with each of the $k$ sub-samples used exactly once as the validation data. For this problem, data was split into 10 sets of size $n/10$, training with 9 subsets and testing on the remaining one subset. This process was repeated ten times while using a different subset for the test each time. The final results were then calculated by taking the mean accuracy of ten tests.

\section{Results}
\label{sec-4}

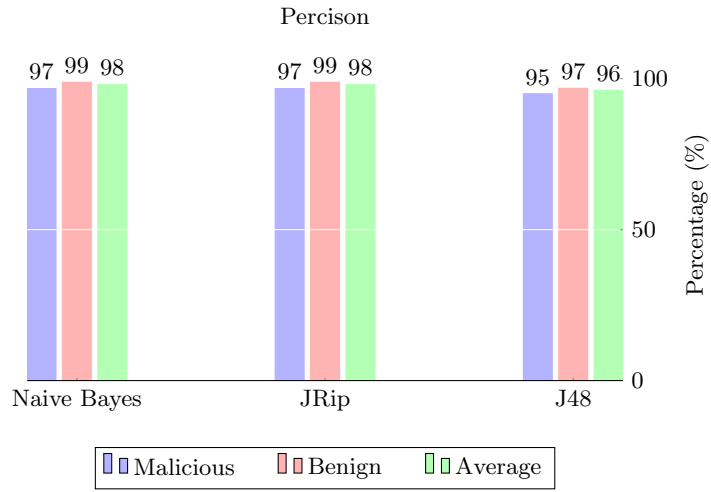
\begin{figure}
	\begin{tikzpicture}
	\centering
	\begin{axis}[
	ybar, axis on top,
	title={Percison},
	height=6cm, width=9.5cm,
	bar width=0.4cm,
	ymajorgrids, tick align=inside,
	major grid style={draw=white},
	enlarge y limits={value=.1,upper},
	ymin=0, ymax=100,
	axis x line*=bottom,
	axis y line*=right,
	y axis line style={opacity=0},
	tickwidth=1pt,
	enlarge x limits=true,
	legend style={
		at={(0.5,-0.2)},
		anchor=north,
		legend columns= -1,
		/tikz/every even column/.append style={column sep=0.5cm}
	},
	ylabel={Percentage (\%)},
	symbolic x coords={
		Naive Bayes,JRip,J48,},
	xtick=data,
	nodes near coords={
		\pgfmathprintnumber[precision=0]{\pgfplotspointmeta}
	}
	]
	\addplot [draw=none, fill=blue!30] coordinates {
		(Naive Bayes,96.9)
		(JRip, 96.9)
		(J48,95.2)
	};
	\addplot [draw=none,fill=red!30] coordinates {
		(Naive Bayes,99)
		(JRip, 99)
		(J48,97)
	};
	\addplot [draw=none, fill=green!30] coordinates {
		(Naive Bayes,98.2)
		(JRip, 98.2)
		(J48,96.3)
	};
	\legend{Malicious,Benign,Average}
	\end{axis}
	\end{tikzpicture}
	
	\caption{Precision of different classifiers}
	\label{precision}
\end{figure}

Figure \ref{precision} shows the comparison of each classifier for malicious, benign, and average instance by using precision metric. We observed that NB scored 97\% Malicious, 99\% Benign, and 98\% Average whereas JRip scored 97\% Malicious, 99\% Benign and 98\% Average. Finally, J48 scored 95\% Malicious, 97\% Benign and 96\% Average. Among all the three classifiers, the J48 scored the lowest with the average score of 96\%. Naive Bayes  and JRip have scored the highest in the tests, with similar results of average being 98\%. Therefore, NB and JRip classifiers are used in the following analysis.

\subsection{Metrics}

\subsubsection{Confusion matrix}
\begin{figure}
\centering
\includegraphics[width=.5\textwidth]{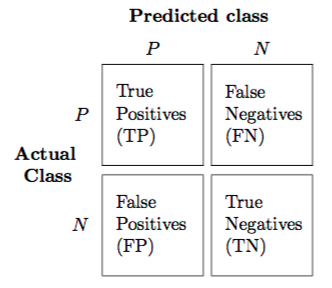}
\caption{Confusion matrix}\label{cm}
\end{figure}

The confusion matrix summarizes the performance of classification model. True Positive (TP), False Negative (FN), False Positive (FP), and True Negative (TN) are elements of confusion matrix as shown in Figure \ref{cm}. Columns represent the predicted class while rows represent the actual class. Higher values in the main diagonal reflect better accuracy in the classification. 

\subsubsection{True positive rate} A true positive rate is the proportion of positives that are correctly identified by classifier. The TP rate is defined as follows.

\begin{equation}
TP Rate=\frac{TP}{TP+FN}
\end{equation}

\subsubsection{False positive rate} A False Positive rate is the proportion of the outcome that is incorrectly predicted as yes (or positive)
when it is actually no (negative).The FP rate is defined as follows.

\begin{equation}
FP Rate=\frac{FP}{FP+TN}
\end{equation}

\subsubsection{Precision} precision is the fraction of relevant instances among the retrieved instances.

\begin{equation}
Precision =\frac{TP}{TP+FP}
\end{equation}

\subsubsection{Recall} Recall is the fraction of relevant instances among the retrieved instances.

\begin{equation}
Recall =\frac{TP}{TP+FN}
\end{equation}

\subsubsection{F-measure}The F-measure is defined as a harmonic mean of precision and recall.

\begin{equation}
Precision =\frac{2  x  Precision  x  Recall}{Recall + Precision}
\end{equation}

\subsubsection{ Matthews correlation coefficient (MCC)} MCC ranges from −1.0 (worst) to 1.0 (best) and is defined as follows.
\begin{equation}
MCC = \frac{( TP x TN) - (FP x FN) }  {\sqrt{ (TP +FP)     (TP+FN)    (TN+FP)     (TN+FN)}}
\end{equation}
\subsection{Naive Bayes}
\begin{table}
\caption{Naive Bayes classifier results}
\label{resultNB}
\centering
\setlength\tabcolsep{2.5em}
\begin{tabular}{llll}
\hline
          & Malware      & Benign       & Average     \\  \hline \\[-2ex]
TPR       & 98.40\%      & 98\%         & 98.20\%     \\
FPR       & 2\%          & 1.60\%       & 1.70\%      \\
Precision & 96.90\%      & 99\%         & 98.20\%     \\
Recall    & 98.40\%      & 98\%         & 98.20\%     \\
F-Measure & 97.60\%      & 98.50\%      & 98.20\%     \\
MCC       & 96.20\%      & 96.20\%      & 96.20\%     \\
ROC Area  & 98.70\%      & 99.50\%      & 99.20\%     \\
PRC Area  & 96\%         & 99.70\%      & 98.30\%     \\ \hline
\end{tabular}
\end{table}
The results of the NB classifier are shown in Table \ref{resultNB}. The average score of TP is 98.20\%, which indicates that the attributes have been correctly identified. The  FP averaged 1.70\%, which indicates that the result is scoring low on the error scale of the attributes. Therefore, the results can be identified as viable and true in this test. Table \ref{CM NB} shows the confusion matrix containing the details of the predicted and actual classes done by the NB classifier. Using these numbers we can calculate the TP and FP rates.

\begin{table}
\caption{Confusion Matrix Naive Bayes}
\label{CM NB}
\centering
\setlength\tabcolsep{1.0em}
\begin{tabular}{lclclcl}
\hline
&\textbf{a=Malicious} & \textbf{b=Benign} \\ \hline  \\[-2ex]
\textbf{a=Malicious}&61                   & 2                 \\
\textbf{b=Benign}&1                    & 100               \\ \hline
\end{tabular}
\end{table}

Applying formula 1 and 2 to the confusion matrix of NB, we get the following results.

\begin{equation}
TP Rate=\frac{61}{61+2}=0.968
\end{equation}
\begin{equation}
FP Rate=\frac{1}{1+100}=0.009
\end{equation}

\subsection{JRIP}
Table \ref{resultJRIP} shows that average TP of 98.20\%, which indicates that JRip is able to correctly classify the URLs. The FP score is 1.70\%, which indicates the classification had a low number of errors.

\begin{table}
\caption{JRip classifier results}
\label{resultJRIP}
\centering
\setlength\tabcolsep{2.5em}
\begin{tabular}{llll}
\hline
          & Malware      & Benign       & Average     \\  \hline \\[-2ex]
TPR       & 98.40\%    & 98\%      & 98.20\%   \\
FPR       & 2\%        & 1.60\%    & 1.70\%    \\
Precision & 90.60\%    & 99\%      & 98.20\%   \\
Recall    & 98.40\%    & 98\%      & 98.20\%   \\
F-Measure & 97.60\%    & 98.50\%   & 98.20\%   \\
MCC       & 96.20\%    & 96.20\%   & 96.20\%   \\
ROC Area  & 97.20\%    & 97.20\%   & 97.20\%   \\
PRC Area  & 92.80\%    & 98\%      & 96\%      \\ \hline
\end{tabular}
\end{table}

In Table \ref{CMJRIP}  the confusion matrix is presented, which contains the details about the predicted and actual classification done by the JRip classifier. The count of TP is 62, FN is 1, and FP is 1 whereas TN is equal to 100. Using these numbers we can calculate the TP rate and FP rate.
\begin{table}
\caption{Confusion Matrix JRip}
\label{CMJRIP}
\centering
\setlength\tabcolsep{1.0em}
\begin{tabular}{lclclcl}
\hline
&\textbf{a=Malicious} & \textbf{b=Benign} \\ \hline  \\[-2ex]
\textbf{a=Malicious}&62                   & 1                 \\
\textbf{b=Benign}&1                    & 100               \\ \hline
\end{tabular}
\end{table}

\begin{equation}
TP Rate=\frac{62}{62+1}=0.984
\end{equation}
\begin{equation}
FP Rate=\frac{1}{1+100}=0.009
\end{equation}

\subsection{J48}
From Table \ref{resultJ48}, we can deduce that the average TP is 96.30\%, which indicates that most of the URLs are correctly classified. The FP score is 4.10\%, which indicates that the classification had a low number of errors.

\begin{table}
\caption{J48 classifier results}
\label{resultJ48}
\centering
\setlength\tabcolsep{2.5em}
\begin{tabular}{llll}
\hline
          & Malware      & Benign       & Average     \\  \hline \\[-2ex]

TPR       & 95.20\%   & 97\%      & 96.30\%   \\
FPR       & 3\%       & 4.80\%    & 4.10\%    \\
Precision & 95.20\%   & 97\%      & 96.30\%   \\
Recall    & 95.20\%   & 97\%      & 96.30\%   \\
F-Measure & 95.20\%   & 97\%      & 96.30\%   \\
MCC       & 92.30\%   & 92.30\%   & 92.30\%   \\
ROC Area  & 95.60\%   & 95.60\%   & 95.60\%   \\
PRC Area  & 91.60\%   & 96.10\%   & 94.40\%   \\ \hline
\end{tabular}
\end{table}

Table \ref{CMJ48} shows the confusion matrix, which contains the details about the predicted and actual classification done by the J48 classifier. Using these numbers we can calculate the TP rate and FP rate.

\begin{table}
\caption{Confusion Matrix J48}
\label{CMJ48}
\centering
\setlength\tabcolsep{1.0em}
\begin{tabular}{lclclcl}
\hline
&\textbf{a=Malicious} & \textbf{b=Benign} \\ \hline  \\[-2ex]
\textbf{a=Malicious}&60                  & 3                 \\
\textbf{b=Benign}&3                   & 98              \\ \hline
\end{tabular}
\end{table}

\begin{equation}
TP Rate=\frac{60}{60+3}=0.952
\end{equation}
\begin{equation}
FP Rate=\frac{3}{3+98}=0.029
\end{equation}

\section{Conclusions}
\label{sec-5}
In this paper, we proposed an approach to filter benign and malicious websites. The URL based analysis is performed that helped by removing the runtime latency and delay of loading the websites. Furthermore, the proposed design protects the users from attacks induced by browser vulnerabilities. The proposed approach can be applied via a blacklisting content and system-based evaluation of site content and behavior of the site. By selecting the right features and algorithms, our system has achieved 98\% accuracy in detecting and classifying the malicious URLs.
The limitation of the work include the small dataset, number of classifiers used and actual real time testing. Future work would include creating a browser plugin and testing the system with real data, using a much larger dataset and investigating deep learning methods.

\section*{Acknowledgment}
This research was supported, in part, by Zayed University Research Office, Research Incentives Grant \# R18054.


\bibliographystyle{splncs03}
\bibliography{example}

\end{document}